\documentclass[aps,pre,twocolumn,groupedaddress]{revtex4}
\usepackage{graphicx}
\usepackage{amsmath}

\begin{document}
\title{Mutual Information and the Edge of Chaos in Reservoir Computers}
\author{T. L. Carroll}
\email{Thomas.Carroll@nrl.navy.mil}
\affiliation{US Naval Research Lab, Washington, DC 20375}

\date{\today}

\begin{abstract}
A reservoir computer is a dynamical system that may be used to perform computations. A reservoir computer usually consists of a set of nonlinear nodes coupled together in a network so that there are feedback paths. Training the reservoir computer consists of inputing a signal of interest and fitting the time series signals of the reservoir computer nodes to a training signal that is related to the input signal. It is believed that dynamical systems function most efficiently as computers at the "edge of chaos", the point at which the largest Lyapunov exponent of the dynamical system transitions from negative to positive. In this work I simulate several different reservoir computers and ask if the best performance really does come at this edge of chaos. I find that while it is possible to get optimum performance at the edge of chaos, there may also be parameter values where the edge of chaos regime produces poor performance. This ambiguous parameter dependance has implications for building reservoir computers from analog physical systems, where the parameter range is restricted.

\end{abstract}

\maketitle

{\bf 
A reservoir computer is a type of recurrent neural network. Because of the recurrence, reservoir computers are dynamical systems, so the methods of nonlinear dynamics should be useful for understanding reservoir computers. Typically a reservoir computer is built by connecting a number of nonlinear nodes into a network in which there are feedback paths. The main requirement for the nodes and the network is that the network is stable, that is it should settle to a stable fixed point. The nodes are then driven by some signal to be analyzed. The individual nodes produce time series signals that are influenced by the input signal; if driven multiple times by the same input signal, the nodes should produce repeatable outputs. The node output signals are combined in a linear combination to fit a training signal. The coefficients for this fit contain information about the relation of the training signal to the input signal, and the performance of the reservoir computer is measured by the error between the training signal and the fit signal.

The theory of computation with dynamical systems states that as a low dimensional dynamical system follows a route to chaos, the complexity of the dynamical system increases, with the greatest complexity coming at the edge of chaos. It is at this point where the dynamical system is said to have its greatest computational capacity. As a result of this edge of chaos concept, it is believed that the optimum performance of a reservoir computer will come when the reservoir computer parameters are tuned to the edge of chaos. In this paper, different reservoir computers are simulated with a range of parameters to see if the edge of chaos is really the best parameter regime to operate a reservoir computer. The results are ambiguous, especially if one wants to create a reservoir computer from analog physical systems.

}
\section{Introduction}
A reservoir computer \cite{jaeger2001,natschlaeger2002} is a nonlinear dynamical system that may be used to perform computations on time series signals. Typically the dynamical system is created by connecting a number of nonlinear nodes in a network that includes paths that form cycles, resulting in feedback. Because of the feedback, a reservoir is part of the class of neural networks known as recurrent neural networks.

 An input signal about which one wants to learn something is used to drive the nonlinear nodes. As with other neural networks, before being used as a computer, a reservoir computer must be trained to extract information from the input signal by training on a signal that contains some useful information about the input signal, but unique to a reservoir computer, the network of internal connections is not altered. Rather, the time series signals produced by the individual nodes are fit to the training signal. As an example, in \cite{lu2017}, the authors sought to reproduce a Lorenz $z$ signal based on a Lorenz $x$ signal, so the $x$ signal was the input and the reservoir computer was trained on the $z$ signal.

Training the reservoir computer consists of creating a linear combination of the time series signals produced by the nonlinear nodes to fit to the training signal. This fit can be as simple as a least squares fit. The training process, therefore, can be much faster than for a neural network. The coefficients for this linear combination are the output from this training.

To compute with a reservoir computer, a signal different but related to the original input signal is used to drive the reservoir. In \cite{lu2017}, the new input signal was the $x$ signal from a Lorenz system started with different initial conditions. The new input signal drives the same network as used for training. A linear combination is then made from the node signals using the coefficients found during the training stage. The signal produced by this linear combination is the output of the computation stage. In \cite{lu2017}, the output of the training stage was the Lorenz $z$ signal corresponding to the new Lorenz $x$ signal.

Because of their simplicity, reservoir computers can be built as analog systems.  Examples of reservoir computers so far include photonic systems \cite{larger2012, van_der_sande2017}, analog circuits \cite{schurmann2004}, mechanical systems \cite{dion2018} and  field programmable gate arrays \cite{canaday2018}. This analog approach means that reservoir computers can potentially be very fast, and yet consume little power, while being small and light. Reservoir computers have been shown to be useful for solving a number of problems, including reconstruction and prediction of chaotic attractors \cite{lu2018,zimmerman2018,antonik2018,lu2017,jaeger2004}, recognizing speech, handwriting or other images \cite{jalavand2018} or controlling robotic systems \cite{lukosevicius2012}.

When reservoir computers are used, the parameters are usually set arbitrarily. If the reservoir computer is built from analog hardware, the parameter choices are limited by the particular implementation, but it would still be useful to know how to get the best performance from the computer. In simulations, there is some conventional wisdom based on experience; the network connecting the nodes should be random, the spectral radius of the network should be less than 1, the network should be operated close to the edge of chaos, and so on \cite{lukosevicius2012}. As a parameter is varied, the edge of chaos is the parameter value at which the network transitions from stable to unstable, usually indicated by the largest Lyapunov exponent of the reservoir network going from below 0 to above 0. These rules are based on simulations of a few types of problems with a few types of nonlinear nodes, so it is not clear if these rules should apply in every situation, and there is evidence that this edge of chaos rule is not universal \cite{mitchell1993,mitchell1993a}. 

In this paper, I vary multiple parameters in different reservoir computers and ask if any measured properties of the signals produced by the reservoir computer correlate with improved performance. By improved performance, I mean that the difference between the linear combination of node time series and the training signal is minimized. I study how the mutual information between the training signal and the signal produced from a linear combination of reservoir signals affects performance, and I also check to see if the best performance comes at the edge of chaos.

\subsection{Edge of Chaos}

The idea that a dynamical system or a celluar automaton has its greatest computational capacity at the edge of chaos was introduced in \cite{packard1988,langton1990,crutchfield1990}. The edge of chaos is a phase transition between an ordered state and a disordered state. Systems in the vicinity of this phase transition exhibit the most complex behavior of any parameter range, and thus have the greatest capacity for computation. Because dynamical systems have the greatest computational capacity at the edge of chaos, it is believed that a reservoir computer will function best at this edge.

In many reservoir computers, depending on the node type, there is no actual chaotic behavior. Instead, when the largest Lyapunov exponent for the reservoir becomes positive, the reservoir network becomes unstable; in simulations, the reservoir signals diverge to positive or negative infinity. It would be more accurate to call the point where the Lyapunov exponent becomes positive the edge of stability, but the term "edge of chaos" is widely used, so that phrase will be used here.

\section{Reservoir Computers}
\label{computers}
\begin{figure}[ht]
\centering
\includegraphics[scale=0.4]{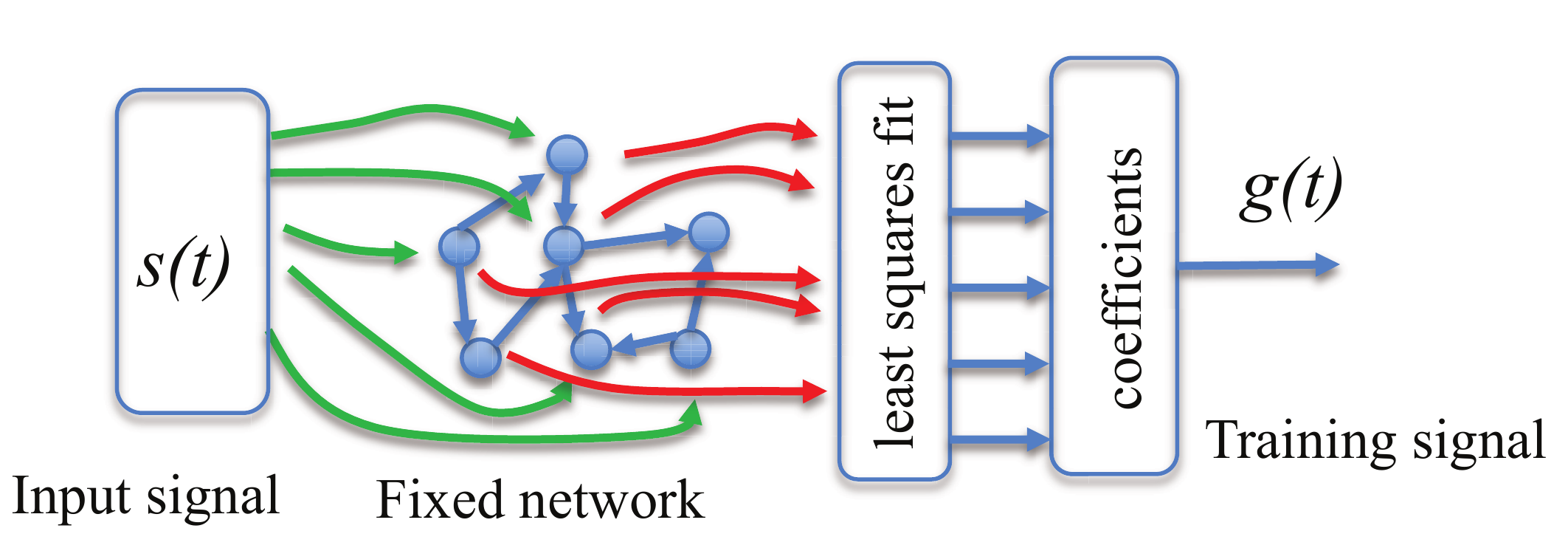} 
  \caption{ \label{reservoir_computer}
Block diagram of a reservoir computer. We have an input signal $s(t)$ that we want to analyze, and a related training signal $g(t)$. When trained, the reservoir computer will estimate $g(t)$ from $s(t)$. In the training phase, the input signal $s(t)$ drives a fixed network of nonlinear nodes, and the time varying signals from the nodes are fit to the training signal $g(t)$ by a least squares fit. The coefficients are the result of the training phase. To use the reservoir computer for computation, a different signal $s'(t)$ is input to the reservoir computer. As an example, \cite{lu2017}, $s(t)$ was a Lorenz $x$ signal, while $s'(t)$ was a Lorenz $x$ signal started with different initial conditions. The time varying node signals that result from $s'(t)$ are multiplied by the coefficients from the training phase to produce the output signal $g'(t)$, which in \cite{lu2017} was a good fit to the Lorenz $z$ signal corresponding to $s'(t)$.}
  \end{figure}

We used a reservoir computer to estimate one time series signal based on a different (but related) time series signal.
Figure \ref{reservoir_computer} is a block diagram of a reservoir computer. There is an input signal $s(t)$ from which the goal is to extract information, and a training signal $g(t)$ which is used to train the reservoir computer. In \cite{lu2017} for example, $s(t)$ was the $x$ signal from a Lorenz chaotic system, while $g(t)$ was the Lorenz $z$ signal. The reservoir computer was trained to estimate the $z$ signal from the $x$ signal.

A reservoir computer may be described by
\begin{equation}
\label{gen_comp}
{r_i}\left( {n + 1} \right) = f\left( {{r_i}\left( n \right) + \sum\limits_{j = 1}^M {{A_{ij}}{r_j}\left( n \right)}  + {w_i}s\left( t \right)} \right)
\end{equation}

where the reservoir computer variables are the $r_i(n), i=1 ... M$ with $M$ the number of nodes, $A$ is an adjacency matrix that described how the different nodes in the network are connected to each other, ${\bf W}=[w_1, w_2, ... w_M]$ describes how the input signal $s(t)$ is coupled into the different nodes, and $f$ is a nonlinear function. 

When the reservoir computer was driven with $s(t)$, the first 2000 time steps were discarded as a transient. The next $N=5000$ time steps from each node were combined in a $N \times (M+1)$ matrix
\begin{equation}
\label{fit_mat}
{\Omega } = \left[ {\begin{array}{*{20}{c}}
{{r_1}\left( 1 \right)}& \ldots &{{r_M}\left( 1 \right)}&1\\
{{r_1}\left( 2 \right)}&{}&{{r_M}\left( 2 \right)}&1\\
 \vdots &{}& \vdots & \vdots \\
{{r_1}\left( N \right)}& \ldots &{{r_M}\left( N \right)}&1
\end{array}} \right]
\end{equation}
The last column of $\Omega $ was set to 1 to account for any constant offset in the fit. The training signal is fit by
\begin{equation}
\label{train_fit_0}
h\left( t \right) = \sum\limits_{j = 1}^M {{c_j}{r_j}\left( t \right)} 
\end{equation}
or

\begin{equation}
\label{train_fit}
{h(t)} ={\Omega } {{\bf C}}
\end{equation}
where ${h(t)} = \left[ {h\left( 1 \right),h\left( 2 \right) \ldots h\left( N \right)} \right]$ is the fit to the training signal ${g(t)} = \left[ {g\left( 1 \right),g\left( 2 \right) \ldots g\left( N \right)} \right]$ and ${{\bf C}} = \left[ {{c_1},{c_2} \ldots {c_N}} \right]$ is the coefficient vector.

The matrix ${ \Omega} $ is decomposed by a singular value decomposition 
\begin{equation}
\label{svd}
{\Omega}   = {{US}}{{{V}}^T}.
\end{equation}
where ${ U}$ is $N \times (M+1)$, ${ S}$ is  $(M+1) \times (M+1)$ with non-negative real numbers on the diagonal and zeros elsewhere, and ${ V}$ is $(M+1) \times (M+1)$.

The pseudo-inverse of ${ \Omega }$ is constructed as a Moore-Penrose pseudo-inverse \cite{penrose1955}
\begin{equation}
\label{pinv}
{ \Omega _{inv}} = {{V}}{{{S}}^{{'}}}{{U}}^T
\end{equation}
where ${{S}}^{{'}}$ is an $(M+1) \times (M+1)$ diagonal matrix constructed from ${ S}$, where the diagonal element $S^{'}_{i,i}=S_{i,i}/(S_{i,i}^2+k^2)$, where $k=1 \times 10^{-5}$ is a small number used for ridge regression \cite{tikhonov1943} to prevent overfitting. There are some guidelines for choosing $k$ \cite{golub1979}, but in this case $k$ is chosen large enough to to keep the coefficients from becoming extremely large but small enough to keep the fitting error from becoming too large.

The fit coefficient vector is then found by
\begin{equation}
\label{fit_coeff}
{{\bf C}} = {{ \Omega } _{inv}}{g(t)}
\end{equation}.

The training error may be computed from
\begin{equation}
\label{train_err}
{\Delta _{RC}} = \frac{{\left\| {  \Omega{{\bf C}} - {g(t)}} \right\|}}{{\left\| {g(t)} \right\|}}
\end{equation}
where $\left\| {} \right\|$ indicates a standard deviation. 

The training error tells us how well the reservoir computer can fit a known training signal, but it doesn't tell us anything we don't already know. To learn new information, we use the reservoir computer in the testing configuration. As an example, suppose the input signal $s(t)$ was an $x$ signal from the Lorenz system, and the training signal $g(t)$ was the corresponding $z$ signal. Fitting the Lorenz $z$ signal trains the reservoir computer to reproduce the Lorenz $z$ signal from the Lorenz $x$ signal.

We may now use as an input signal $s'(t)$ the Lorenz signal $x'$, which comes from the Lorenz system with different initial conditions. We want to get the corresponding $z'$ signal. The matrix of signals from the reservoir is now $\Omega'$. The coefficient vector ${\bf C}$ is the same vector we found in the training stage. The testing error is
\begin{equation}
\label{test_err}
{\Delta _{tx}} = \frac{{\left\| {\Omega '{\bf{C}} - z'} \right\|}}{{\left\| {z'} \right\|}}
\end{equation}
The testing error measures how accurately the reservoir computer actually solves a problem.

\section{The Input Coupling Vector ${\bf W}$}
\label{input}
The coupling vector ${\bf W}=[w_1, w_2, ... w_M]$ describes how the input signal $s(t)$ couples into each of the nodes. I want to change only specified parameters in the reservoir computer, so ${\bf W}$ is kept fixed. It has been found that setting all the elements to +1 or -1 yields a larger reservoir computer testing error than setting the odd elements of ${\bf W}$ to +1 and the even elements of ${\bf W}$ to -1, so the second method (odd=+1, even=-1) was used. This choice was arbitrary, and other choices of ${\bf W}$ could be made.

\section{Network}
\label{network}
The network was kept fixed so that changes in the network would not affect the results of the parameter variation. The effect of different networks on reservoir computer performance was studied in \cite{carroll2019}. The adjacency matrix was initialized to a matrix where 20\% of the network edges were +1 while the others were 0. All the diagonal entries in the adjacency matrix were 0. Of the edges with a value of +1, 50\%  were then flipped to -1. The edges to flip were chosen randomly. The adjacency matrix was then normalized so that the spectral radius $\rho$, defined as the absolute value of the largest real part of the eigenvalues of the matrix, was set to a specified value.

\section{Node Types}

There are no specific requirements on the nodes in a reservoir computer, other than when all nodes are connected into a network, the network should be stable; that is, when not driven, it should settle into a stable fixed point, and when driven, the same input signal should produce repeatable outputs. Several different node types are used here to decrease the chance that the results depend only on the type of node used.

The polynomial reservoir computer is described by
\begin{equation}
\label{res_comp}
\begin{split}
&\frac{{d{r_i}\left( t \right)}}{{dt}} = \\ &\lambda \left[ {{p_1}{r_i}\left( t \right) + {p_2}r_i^2\left( t \right) + {p_3}r_i^3\left( t \right) 
  + \sum\limits_{j = 1}^M {{A_{ij}}{r_j}\left( t \right)}  + {w_i}s\left( t \right)} \right].
  \end{split}
\end{equation}
The $r_i(t)$'s  are node variables, $A$ is an adjacency matrix indicating how the nodes are connected to each other, and ${\bf W}=[w_1, w_2, ... w_M]$ is a vector that describs how the input signal $s(t)$ is coupled to each node. The constant $\lambda$ is a time constant, and there are $M=100$ nodes. This node type will be called the  polynomial  node. For the simulations described here,  $p_2=1$, $p_3=-1$ and $\lambda=1.4$, while $p_1$ varied.  These nonlinear differential equation nodes were chosen because they represent a polynomial, which is a general way to approximate a nonlinear function. The polynomial differential equations were numerically integrated with a 4th order Runge-Kutta integrator with a time step of 0.1 s. 

Reservoir computers with nodes that implement a hyperbolic tangent function are commonly used in the literature. The tanh node computer is described as
\begin{equation}
\label{tanh_comp}
{r_i}\left( {n + 1} \right) = \alpha \tanh \left[ {r{{\left( n \right)}_i} + \sum\limits_{j = 1}^M {{A_{ij}}{r_j}\left( n \right)}  + {w_i}s\left( t \right)} \right].
\end{equation}
Again, $s(t)$ was normalized to have a mean of 0 and a standard deviation of 1. The parameter $\alpha$ could vary

Another node type that was studied was the leaky tanh model from \cite{jaeger2007}
\begin{equation}
\label{umd_comp}
\begin{split}
&{r_i}\left( {n + 1} \right) = \\& \alpha {r_i}\left( n \right) + \left( {1 - \alpha } \right)\tanh \left( {\sum\limits_{j = 1}^M {{A_{ij}}{r_j}\left( n \right)}  + {w_i}s\left( t \right) + 1} \right).
\end{split}
\end{equation}
Tis leaky tanh map was also used in \cite{lu2017,lu2018}.

The leaky tanh node could also be implemented as a flow:
\begin{equation}
\label{umd_flow}
\begin{split}
&\frac{{d{r_i}\left( t \right)}}{{dt}} = \\ & \lambda \left[ {-\alpha {r_i}\left( t \right) + \left( {1 - \alpha } \right)\left[ {\sum\limits_{j = 1}^M {{A_{ij}}{r_j}\left( t \right)}  + {w_i}s\left( t \right) + 1} \right]} \right]
\end{split}
\end{equation}
This system was integrated with a 4'th order Runge-Kutta integrator with a step size of 0.1 s.  The factor $\lambda$ set the overall time scale for the differential equation. Initially, $\lambda$ was set to 1.0.

\section{Input and Training Signals}
The Lorenz system was used to generate input and training signals  \cite{lorenz1963}
\begin{equation}
\label{lorenz}
\begin{array}{l}
\frac{{dx}}{{dt}} = {c_1}y - {c_1}x\\
\frac{{dy}}{{dt}} = x\left( {{c_2} - z} \right) - y\\
\frac{{dz}}{{dt}} = xy - {c_3}z
\end{array}
\end{equation}

with $c_1$=10, $c_2$=28, and $c_3$=8/3. The equations were numerically integrated with a time step of $t_s=0.02$.

\section{Quantities to Measure}
Besides the testing error $\Delta_{tx}$, there were a number of measurements that could be made on the reservoir computer variables. For this work, besides $\Delta_{tx}$, measurements were made of  the maximum Lyapunov exponent for the reservoir $\lambda_{max}$, the mutual information $I[g(t),h(t)]$ between the training signal $g(t)$ and the signal $h(t)$ produced by fitting the reservoir variables to $g(t)$ , and the rank $\Gamma$ of the covariance of reservoir matrix $\Omega$.

The Lyapunov exponent spectrum of the reservoir depended on the input signal $s(t)$, which meant that these were conditional Lyapunov exponents, defined in \cite{pecora1990}.   The original concept of the edge of chaos was developed for autonomous systems, so it is ambiguous for nonautonomous systems if the edge of chaos describes the system with or without the driving signal. We choose to calculate the stability for the driven system, because stability will depend on the driving signal.  Calculation of Lyapunov exponents from a numerical system is well known \cite{eckmann1985}, and the other measurements will be defined here.

\subsection{Entropy}
\label{mi_sect}
To compute the entropy, each signal was transformed into a symbolic time series using the ordinal pattern method \cite{bandt2002}. Each signal was divided into windows of 4 points, and the points within the window were sorted to establish their order; for example, if the points within a window were 0.1, 0.3, -0.1 0.2, the ordering would be 2,4,1,3. Each possible ordering of points in a signal $u(t)$ represented a symbol $\sigma_u(i), i=1 ... N_{su}$, where $N_{su}$ was the number of possible symbols in $u(t)$.  The probabilities $p(\sigma_u(i))$  were found for each  symbol and the entropy was calculated as
\begin{equation}
\label{entropy}
{H_u} - \sum\limits_{i = 1}^{{N_{su}}} {p\left( {{\sigma _u}\left( i \right)} \right)} {\log _2}\left[ {p\left( {{\sigma _u}\left( i \right)} \right)} \right] .
\end{equation}

\subsection{Covariance Rank}
\label{comp_rank}
The individual columns of the reservoir matrix $\Omega$ will be used as a basis to fit the training signal $g(t)$. Among other things, the fit will depend on the number of orthogonal columns in $\Omega$.

Principal component analysis \cite{joliffe2011} states that the eigenvectors of the covariance matrix of $\Omega$, $\Theta=\Omega^T\Omega$, form an uncorrelated basis set. The rank of the covariance matrix tells us the number of uncorrelated vectors. Therefore, we will use the rank of the covariance matrix of $\Omega$,
\begin{equation}
\label{rank}
\Gamma  = {\rm{rank}}\left( {\Omega ^T\Omega } \right)
\end{equation}
to characterize the reservoir matrix $\Omega$. We calculate the rank using the MATLAB rank() function. The maximum covariance rank is equal to the number of nodes, $M=100$.

\section{Varying Parameters}
\label{varpar}
Two parameters were varied for each reservoir type. For each configuration, the testing error $\Delta_{tx}$ was plotted on a contour plot as a function of the the two variables. The maximum Lyapunov exponent for the reservoir, $\lambda_{max}$, the mutual information $I[g(t),h(t)]$ between the training signal $g(t)$ and the signal $h(t)$ produced by fitting the reservoir variables to $g(t)$ , and the rank $\Gamma$ of the covariance of reservoir matrix $\Omega$ were also measured. The testing error was shown on the contour plot and then a line was drawn on the contour plot through the minimum and approximately perpendicular to the contours. The values of the other measurements listed here were plotted for locations in the parameter plane along this line to determine how these quantities compared to the testing error.

\subsection{Polynomial nodes}
The linear parameter $p_1$ and the spectral radius of the network $\rho$ were varied for a reservoir computer whose nodes were described by the polynomial flow of eq. (\ref{res_comp}).

\begin{figure}
\centering
\includegraphics[scale=0.8] {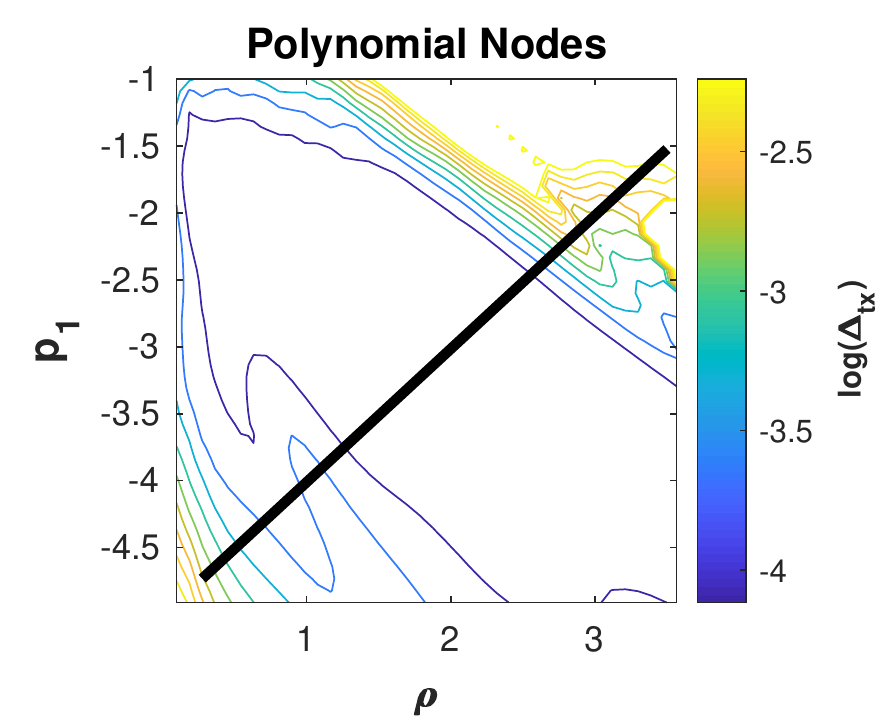} 
  \caption{ \label{nleq_lorenz_2d} Contour plot of the natural log of the reservoir computer testing error $\Delta_{tx}$ as a function of the two parameters $p_1$ from eq. (\ref{res_comp}) and the spectral radius $\rho$ of the reservoir network. The input signal $s(t)$ was a Lorenz $x$ signal while the training signal $g(t)$ was a Lorenz $z$ signal.  The spectral radius $\rho$ is the absolute value of the largest real part of the set of eigenvalues for the network adjacency matrix $A$. Various other measurements for the reservoir computer will be calculated along the black line superimposed on the figure.}
  \end{figure}

Figure \ref{nleq_lorenz_2d} is a contour plot of the testing error $\Delta_{tx}$ for the polynomial flow nodes when the input signal $s(t)$ was the Lorenz $x$ signal (eq. \ref{lorenz})  and the training signal $g(t)$ was the Lorenz $z$ signal.

\begin{figure}
\centering
\includegraphics [scale=0.8]{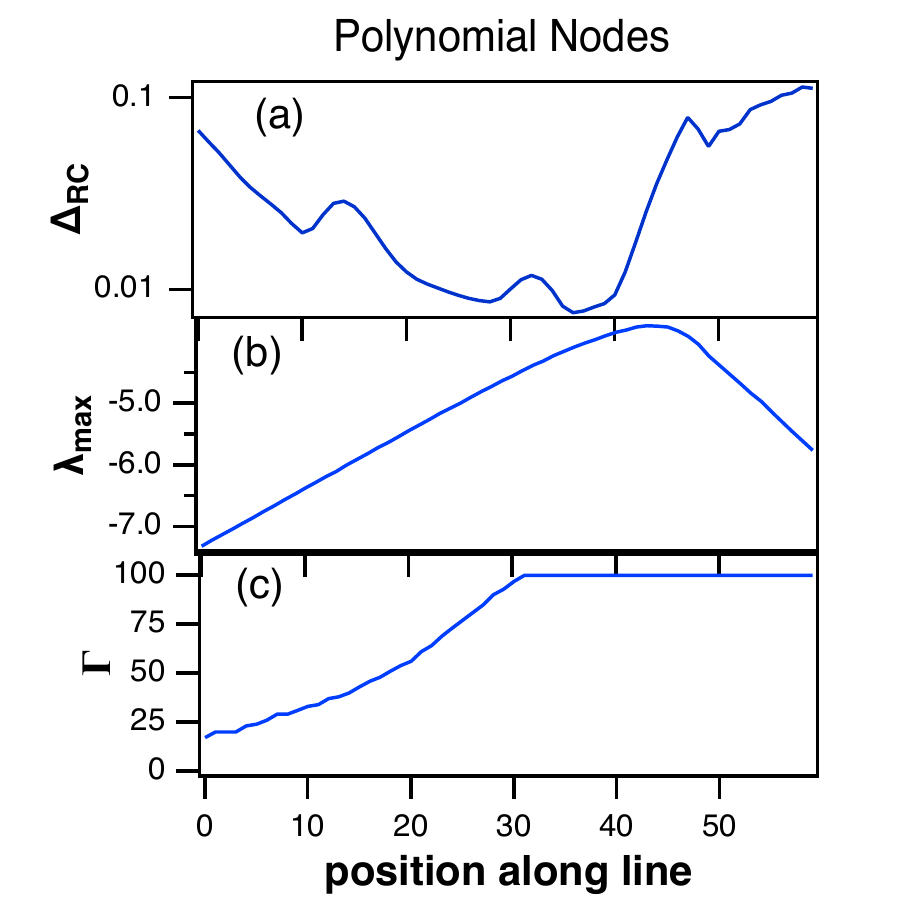} 
  \caption{ \label{lorenz_nleq_line} 
  Plots of different measured quantities along the black line superimposed on fig. \ref{nleq_lorenz_2d}. The input signal $s(t)$ was a Lorenz $x$ signal while the training signal $g(t)$ was a Lorenz $z$ signal.  The left end of the plots corresponds to $[\rho,p_1]=[0.2778, -4.733]$ while the right end of the plots corresponds to $[\rho,p_1]=[3.5, -1.5]$. (a) is the reservoir computer testing error $\Delta_{tx}$,  (b) is the maximum Lyapunov exponent for the reservoir $\lambda_{max}$, and (c) is the rank $\Gamma$ of the covariance of the reservoir computer matrix $\Omega$.}
  \end{figure}

The values of the maximum Lyapunov exponent for the reservoir, $\lambda_{max}$, the mutual information $I[g(t),h(t)]$ between the training signal $g(t)$ and the fit signal $h(t)$ , and the rank $\Gamma$ of the covariance of reservoir matrix $\Omega$ were plotted in fig. \ref{lorenz_nleq_line} for values of $\rho$ and $p_1$ along the line superimposed on fig.  \ref{nleq_lorenz_2d}

In figure \ref{lorenz_nleq_line}, the maximum of the mutual information between the training signal $g(t)$ and reservoir computer fit signal $h(t)$ coincides with the minimum of the testing error $\Delta_{tx}$. The minimum testing error occurs near the maximum of the largest Lyapunov exponent in fig. \ref{lorenz_nleq_line}(c), but the Lyapunov exponent does not cross from less than 0 to greater than 0, so the minimum testing error does not occur at the edge of chaos for these parameters, although it could be that the parameter range plotted does not include the true edge of chaos. The minimum testing error also appears to coincide with the parameter values where the covariance rank $\Gamma$ in fig. \ref{lorenz_nleq_line}(d) saturates at 100.

\subsection{Hyperbolic Tangent nodes}
The hyperbolic tangent is a different nonlinearity than the polynomial nonlinearity used in the previous section, so the reservoir computer behavior may be different. When the reservoir computer nodes were described by a hyperbolic tangent function, as in eq. (\ref{tanh_comp}), the parameters that were varied were the feedback parameter $\alpha$ and the network spectral radius $\rho$.

\begin{figure}[ht]
\centering
\includegraphics [scale=0.8]{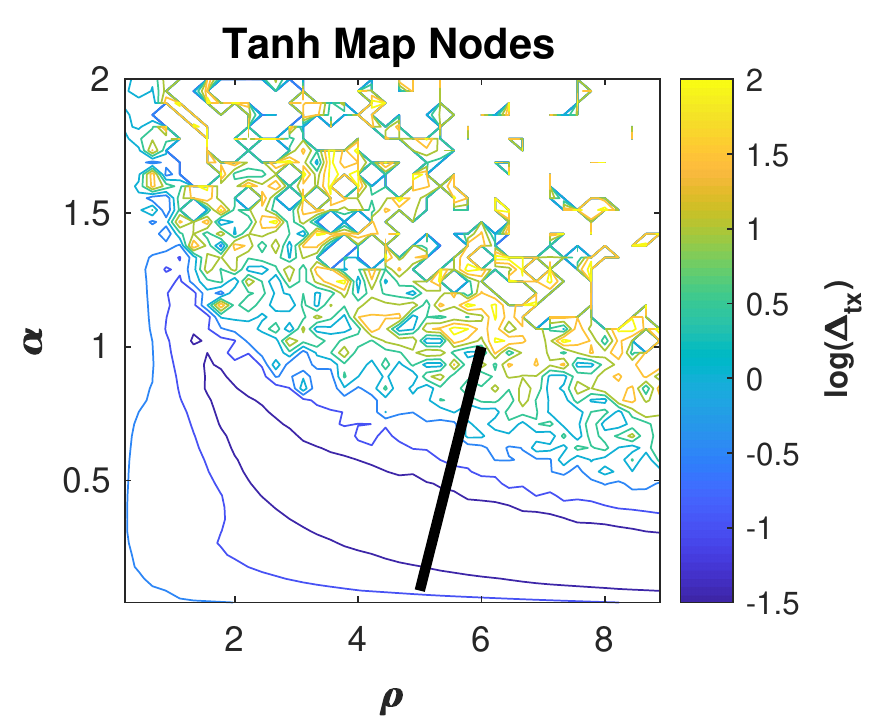} 
  \caption{ \label{tanh_lorenz_2d} Contour plot of the natural log of the reservoir computer testing error $\Delta_{tx}$ as a function of the two parameters $\alpha$ from eq. (\ref{tanh_comp}) and the spectral radius $\rho$ of the reservoir network, for a reservoir computer with the hyperbolic tangent nodes of eq. (\ref{tanh_comp}). The input signal $s(t)$ was a Lorenz $x$ signal while the training signal $g(t)$ was a Lorenz $z$ signal.  The spectral radius is the absolute value of the largest real part of the set of eigenvalues for the network adjacency matrix $A$. Various other parameters for the reservoir computer will be calculated along the black line superimposed on the figure.}
  \end{figure}

Figure \ref{tanh_lorenz_2d} is a contour plot of the testing error $\Delta_{tx}$ for the hyperbolic tangent (tanh) nodes when the input signal $s(t)$ was the Lorenz $x$ signal (eq. \ref{lorenz})  and the training signal $g(t)$ was the Lorenz $z$ signal.

 The contour plot of $\Delta_{tx}$ for the hyperbolic tangent nodes in fig. \ref{tanh_lorenz_2d} is very different than the equivalent plot for the polynomial flow nodes (fig. \ref{nleq_lorenz_2d}). In the upper right section of fig. \ref{tanh_lorenz_2d}, many areas have no contours because the reservoir becomes unstable for some parameter combinations in this region.
 
\begin{figure}[ht]
\centering
\includegraphics [scale=0.8]{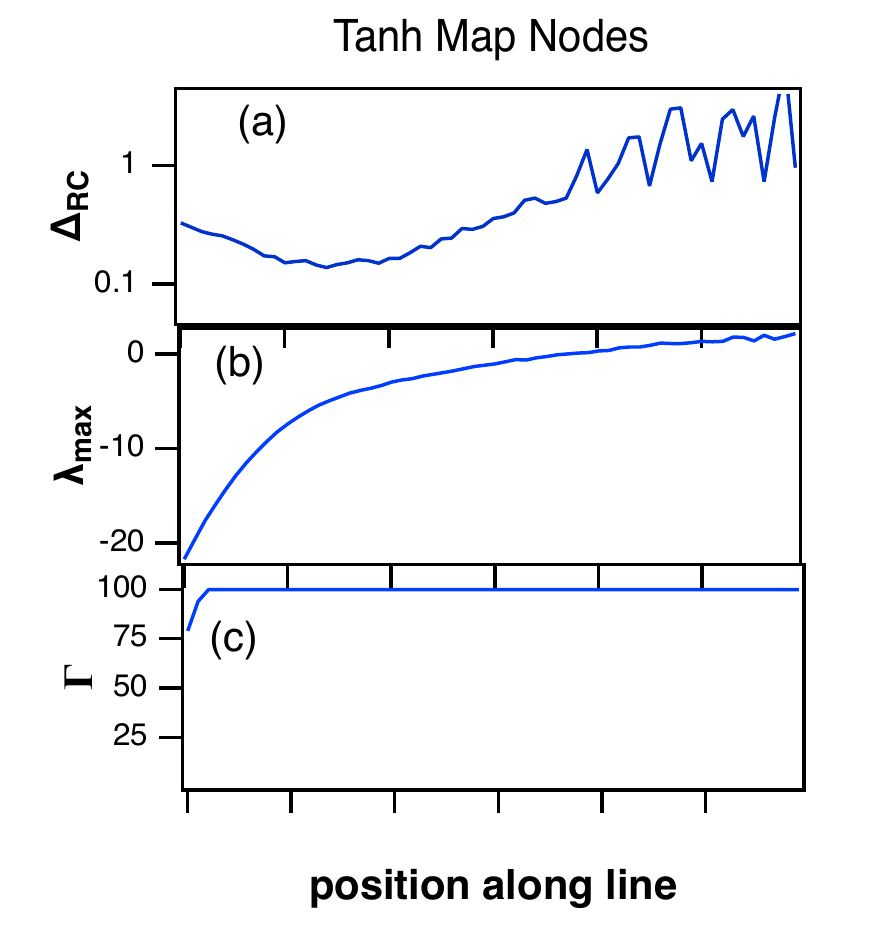} 
  \caption{ \label{tanh_lorenz_line} 
  Plots of different measured quantities along the black line superimposed on fig. \ref{tanh_lorenz_2d}, for hyperbolic tangent nodes of eq. (\ref{tanh_comp}). The input signal $s(t)$ was a Lorenz $x$ signal while the training signal $g(t)$ was a Lorenz $z$ signal.  The left end of the plots corresponds to $[\rho,\alpha]=[5, 0.089]$ while the right end of the plots corresponds to $[\rho,\alpha]=[6, 1]$. (a) is the reservoir computer testing error $\Delta_{tx}$,  (b) is the maximum Lyapunov exponent for the reservoir $\lambda_{max}$, and (c) is the rank $\Gamma$ of the covariance of the reservoir computer matrix $\Omega$.}
  \end{figure}

 Figure \ref{tanh_lorenz_line} shows the values of the other measurements for this system calculated along the black line superimposed on figure \ref{tanh_lorenz_2d}.
 
In figure \ref{tanh_lorenz_line}, the minimum for $\Delta_{tx}$ occurs for the same parameter values as the maximum in the mutual information  $I[g(t), h(t)]$, as it did in figure \ref{lorenz_nleq_line}. The covariance rank $\Gamma$ saturates at its maximum value to the left of the minimum of $\Delta_{tx}$, so the rank is at its maximum value when $\Delta{RC}$ is minimized, but it is not clear if the minimum of $\Delta_{tx}$ is associated with the saturation of the rank.

The minimum in $\Delta_{tx}$ occurs near the maximum of the largest Lyapunov exponent $\lambda_{max}$, but the Lyapunov exponent does not cross 0 along this line, so the minimum of $\Delta_{tx}$ does not occur at the edge of chaos for these parameters. Once again, it is possible that these parameters do not extend to the true edge of chaos.

\subsection{Leaky tanh map}
This section describes a reservoir computer whose nodes were described by a leaky hyperbolic tangent function as in eq. (\ref{umd_comp}). The parameters that were varied were the feedback parameter $\alpha$ and the network spectral radius $\rho$. This map was described in \cite{jaeger2007} and was also used in \cite{lu2017,lu2018}.

\begin{figure}[ht]
\centering
\includegraphics[scale=0.8]{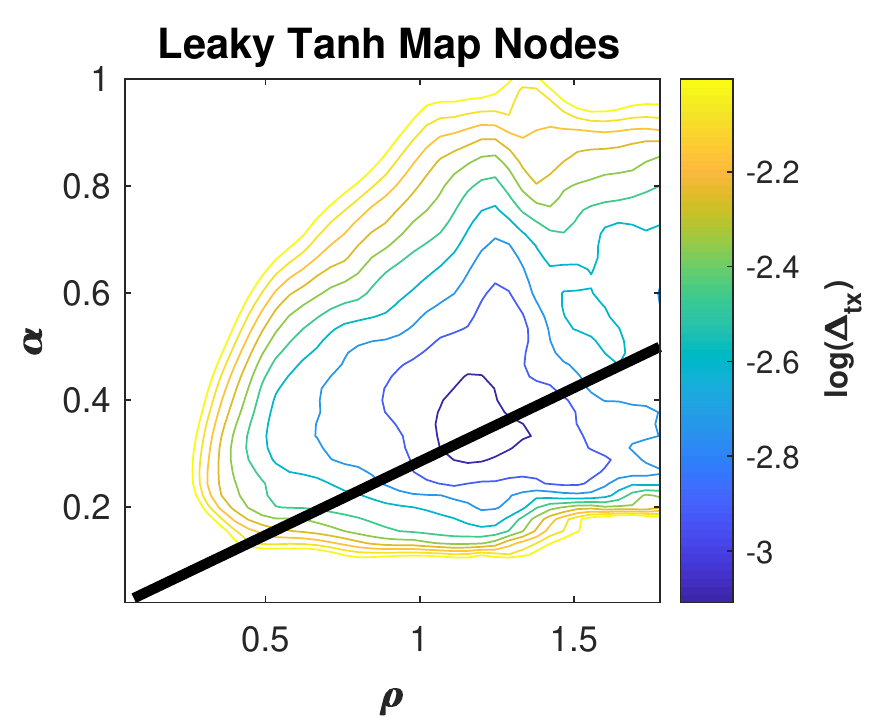} 
  \caption{ \label{umd_lorenz_2d} Contour plot of the natural log of the reservoir computer testing error $\Delta_{tx}$ as a function of the two parameters $\alpha$ from eq. (\ref{umd_comp}) and the spectral radius $\rho$ of the reservoir network, for the leaky tanh map of eq. (\ref{umd_comp}). The input signal $s(t)$ was a Lorenz $x$ signal while the training signal $g(t)$ was a Lorenz $z$ signal.  The spectral radius is the absolute value of the largest real part of the set of eigenvalues for the network adjacency matrix $A$. Various other parameters for the reservoir computer will be calculated along the black line superimposed on the figure.}
  \end{figure} 
  
The reservoir computer built from leaky tanh map nodes does not become unstable for the range of parameters in figure \ref{umd_lorenz_2d}. Equation (\ref{umd_comp}) shows that the range of the parameter $\alpha$ can only be varied between 0 to 1. Figure \ref{umd_lorenz_2d} is a contour plot of $\Delta_{tx}$ for the leaky tanh map.  
  
\begin{figure}[ht]
\centering
\includegraphics[scale=0.8] {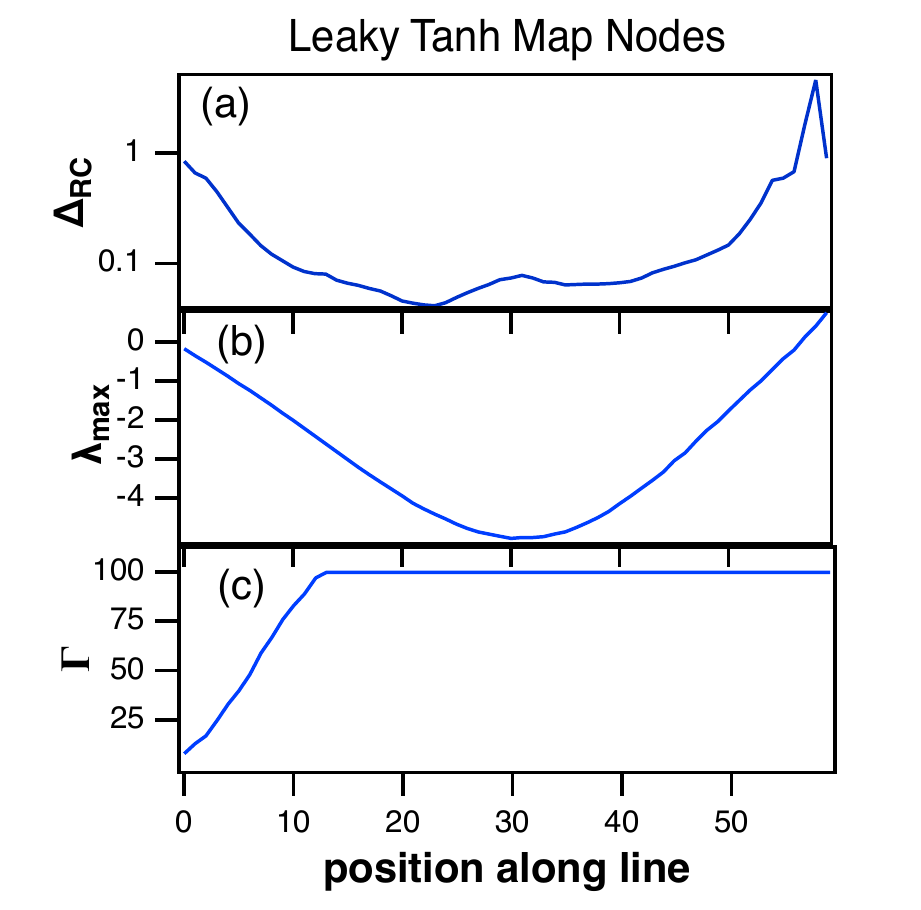} 
  \caption{ \label{umd_lorenz_line} 
  Plots of different measured quantities along the black line superimposed on fig. \ref{umd_lorenz_2d}. The input signal $s(t)$ was a Lorenz $x$ signal while the training signal $g(t)$ was a Lorenz $z$ signal. The nodes were described by the leaky tanh map of eq. (\ref{umd_comp}). The left end of the plots corresponds to $[\rho,\alpha]=[0.01,0.01]$ while the right end of the plots corresponds to $[\rho,\alpha]=[3,1]$. (a) is the reservoir computer testing error $\Delta_{tx}$,  (b) is the maximum Lyapunov exponent for the reservoir $\lambda_{max}$, and (c) is the rank $\Gamma$ of the covariance of the reservoir computer matrix $\Omega$.}
  \end{figure}    
  
  Figure \ref{umd_lorenz_line} shows several different measurements for the leaky tanh reservoir computer driven by the Lorenz $x$ signal when $\alpha$ and $\rho$ were varied along the black line plotted on figure \ref{umd_lorenz_2d}. Once again, the minimum testing error $\Delta_{tx}$ occurs where the mutual information $I[g(t),h(t)]$ between $g(t)$ and $h(t)$ is at a maximum. The minimum of $\Delta_{tx}$ occurs near where the covariance rank $\Gamma$ saturates at 100, but the minimum of $\Delta_{tx}$ and the saturation point of $\Gamma$ are far enough apart that it is not possible to say if they are related. The minimum of $\Delta_{tx}$ does occur when $\Gamma$ is at its maximum value, so the minimum of $\Delta_{tx}$ in all the measurements plotted in this paper does occur when $\Gamma$ is at its maximum value.
  
The plot of $\lambda_{max}$ in figure \ref{umd_lorenz_line} shows that not only does the minimum of $\Delta_{tx}$ not occur at the edge of chaos for this range of variables, the minimum of $\Delta_{tx}$ is near the minimum of $\lambda_{max}$.

\subsection{Leaky tanh flow}
\begin{figure}[ht]
\centering
\includegraphics[scale=0.8]{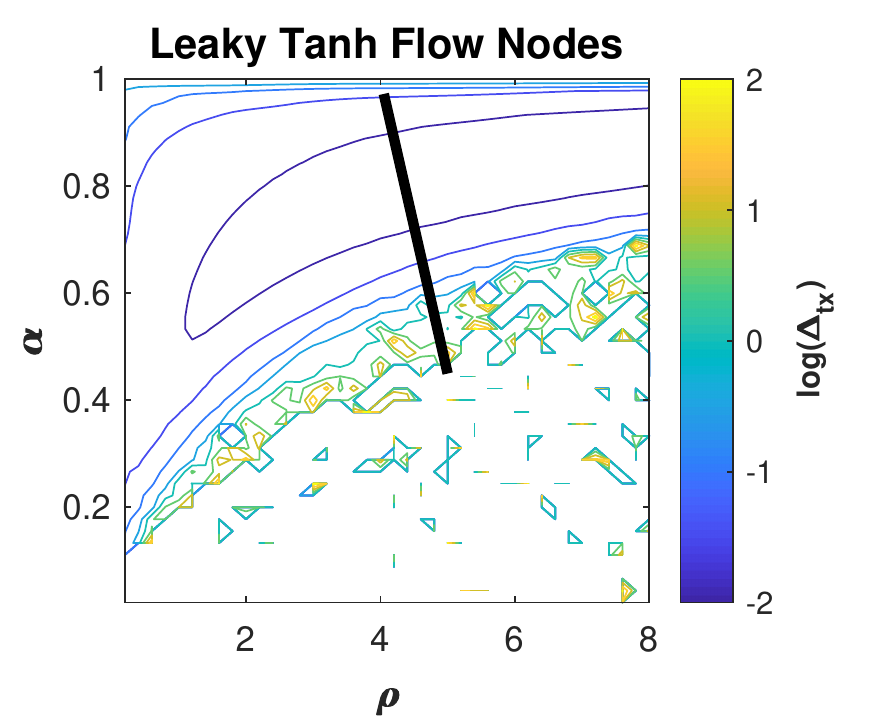} 
  \caption{ \label{lorenz_tanh_flow_2d} Contour plot of the natural log of the reservoir computer testing error $\Delta_{tx}$ as a function of the two parameters $\alpha$ from eq. (\ref{umd_flow}) and the spectral radius $\rho$ of the reservoir network, for the leaky tanh flow of eq. (\ref{umd_flow}). The input signal $s(t)$ was a Lorenz $x$ signal while the training signal $g(t)$ was a Lorenz $z$ signal.  The spectral radius is the absolute value of the largest real part of the set of eigenvalues for the network adjacency matrix $A$. Various other parameters for the reservoir computer will be calculated along the black line superimposed on the figure.}
  \end{figure}

The leaky tanh flow equation (eq. \ref{umd_flow}) looks very similar to the leaky tanh map (eq. \ref{umd_comp}), but the performance as part of a reservoir computer is different. Figure \ref{lorenz_tanh_flow_2d} is a contour plot of the reservoir computer testing error $\Delta_{tx}$ as a function of the feedback parameter $\alpha$ and the network spectral radius $\rho$. 

\begin{figure}[ht]
\centering
\includegraphics[scale=0.8] {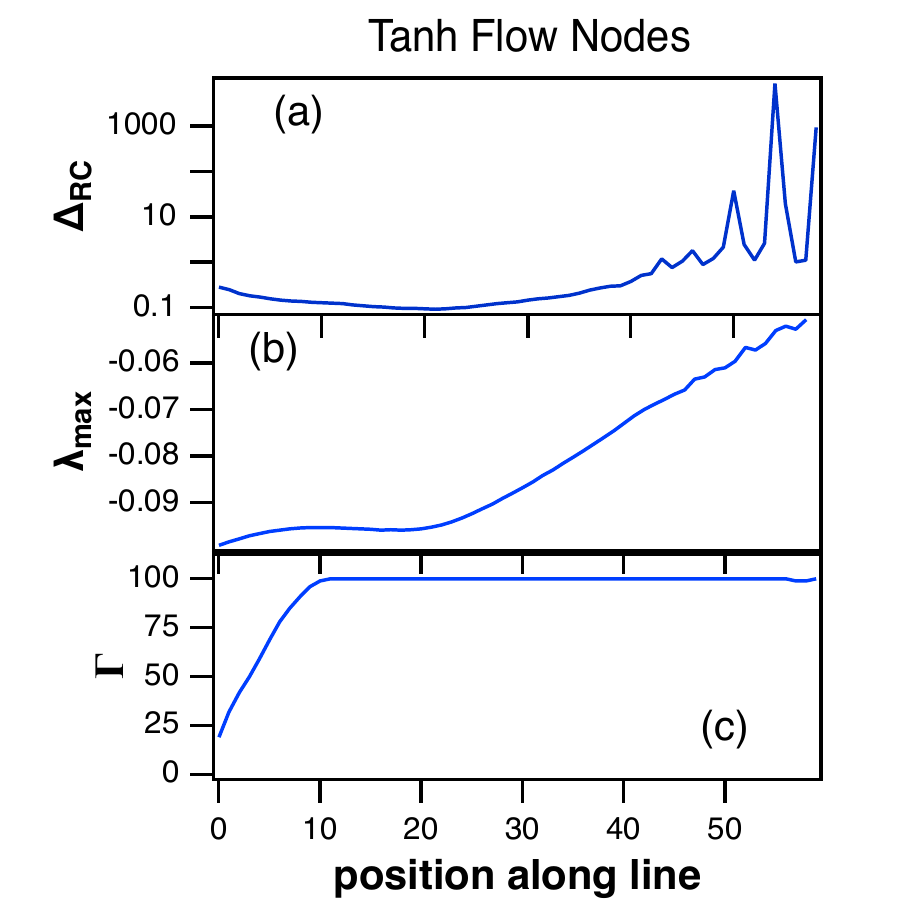} 
  \caption{ \label{tanh_flow_lorenz_line} 
  Plots of different measured quantities along the black line superimposed on fig. \ref{lorenz_tanh_flow_2d}. The input signal $s(t)$ was a Lorenz $x$ signal while the training signal $g(t)$ was a Lorenz $z$ signal. The nodes were described by the leaky tanh flow of eq. (\ref{umd_flow}). The left end of the plots corresponds to $[\rho,\alpha]=[2, 0.01]$ while the right end of the plots corresponds to $[\rho,\alpha]=[4.3,0.4]$. (a) is the reservoir computer testing error $\Delta_{tx}$,  (b) is the maximum Lyapunov exponent for the reservoir $\lambda_{max}$, and (c) is the rank $\Gamma$ of the covariance of the reservoir computer matrix $\Omega$.}
  \end{figure}

The leaky tanh flow network was unstable for a large range of the parameters in fig. \ref{lorenz_tanh_flow_2d}. The upper right part of this figure is mostly blank because the reservoir was unstable for these parameter values. The minimum of $\Delta_{tx}$ in figure \ref{tanh_flow_lorenz_line} was seen for the feedback parameter $\alpha=0$. Figure \ref{tanh_flow_lorenz_line} shows several measurements calculated along the line superimposed on figure \ref{lorenz_tanh_flow_2d}.

As with all the other node types, in figure \ref{tanh_flow_lorenz_line} the minimum testing error $\Delta_{tx}$ occurs where the mutual information $I[g(t),h(t)]$ between $g(t)$ and $h(t)$ is at a maximum. The covariance rank $\Gamma$ is saturated at 100 for most of the parameter range in figure \ref{tanh_flow_lorenz_line}, so it gives no useful information. The maximum Lyapunov exponent is near its minimum value for the minimum in $\Delta_{tx}$, so this plot indicates that for these parameters, the optimum reservoir computer performance does not come at the edge of chaos.
  
It is possible that the edge of chaos really is the best parameter setting to operate all these reservoir computers, but the ranges of parameters studied in these examples is too limited to see this. To further explore different parameter regimes, the reservoir computers were simulated with many different randomly selected parameter combinations.

\section{Random Parameters}

\subsection{Edge of Chaos?}
\label{randpar}
The minimum of the testing error $\Delta_{tx}$ did not occur at the edge of chaos, where the largest Lyapunov exponent of the reservoir goes from negative to positive, for all the combinations of node type and parameter range above. The parameter ranges used in those simulations were restricted. If there are no restrictions on the parameters, is the concept of the edge of chaos useful to describe the optimum performance for a reservoir computer?

To answer this question, reservoir computers with random parameters were simulated. For each combination of input signal and node type, 6000 random combinations of the parameter settings used in creating the contour plots were chosen. The same adjacency matrix $A$ was used for all examples to avoid complications caused by different network configurations. The adjacency matrix was the same as the one described in section \ref{network} above.

\begin{figure}[ht]
\centering
\includegraphics[scale=0.8] {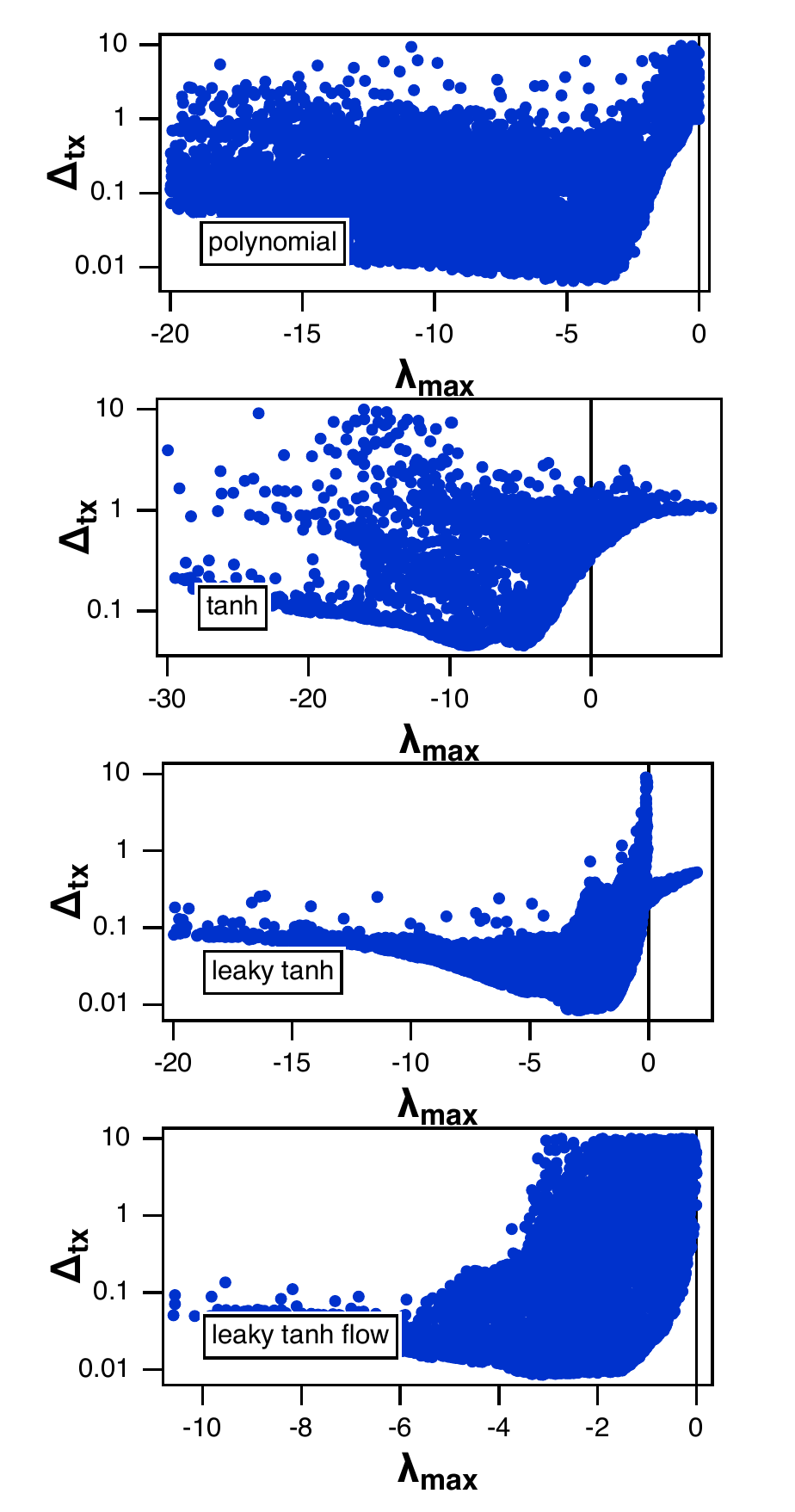} 
  \caption{ \label{random_errors} The blue dots are the reservoir computer testing error $\Delta_{tx}$ as a function of the maximum Lyapunov exponent for the reservoir, $\lambda_{max}$. The input signal $s(t)$ was the Lorenz $x$ signal, while the training signal $g(t)$ was the Lorenz $z$ signal. The node types were the polynomial nodes, the hyperbolic tangent (tahn) nodes, the leaky hyperbolic tangent (leaky tanh) nodes of the leaky hyperbolic tangent flow (leaky tanh flow) nodes.
  }
  \end{figure}

Figure \ref{random_errors} shows the testing error $\Delta_{tx}$ as a function of the maximum Lyapunov exponent for the reservoir, $\lambda_{max}$, when the input signal $s(t)$ was the Lorenz $x$ signal and the training signal $g(t)$ was the Lorenz $z$ signal.Each blue dot in figure \ref{random_errors} shows the maximum Lyapunov exponent and testing error for one realization of the random parameters.

 For the polynomial nodes, each time the reservoir was driven, the spectral radius for the network, $\rho$ was randomly chosen from a uniform distribution between 0 and 1, the parameter $p_1$ was randomly chosen between -4 and 1, $p_2=1$, $p_3$ was randomly chosen between -2 and 0 and the time factor $\lambda$ was randomly chosen between 0 and 10.

For the hyperbolic tangent nodes in figure \ref{random_errors}, for each point, the network spectral radius $\rho$ was chosen from a uniform random distribution between 0 and 10, while the constant $\alpha$ was chosen between 0 and 3. 

For the leaky hyperbolic tangent nodes in figure \ref{random_errors}, the network spectral radius $\rho$ was chosen from a uniform random distribution between 0 and 1, while the constant $\alpha$ was chosen randomly between 0 and 4.

For the leaky hyperbolic tangent flow nodes in figure \ref{random_errors}, the network spectral radius $\rho$ was chosen from a uniform random distribution between 0 and 1,the constant $\alpha$ was chosen randomly between 0 and 4 and the time scale factor $\lambda$ was between 0 and 10.

The simulations of the four different nodes types in figure \ref{random_errors} do not give a definitive answer to the question of whether it is best to operate a reservoir computer on the edge of chaos. One question is how close to the point where the reservoir network becomes unstable does one have to be to say that the reservoir is on the edge of chaos? The minimum testing error for the hyperbolic tangent and the leaky hyperbolic tangent maps in figure \ref{random_errors} is closer to the unstable point than for the polynomial flow or the leaky hyperbolic tangent flow.

There is the additional problem that the lowest testing error may occur near the edge of chaos, but figure \ref{random_errors} shows that some of the largest testing errors also are seen in this region. The random parameter variations show that for some node types, having the reservoir computer parameters near the edge of chaos is necessary but not sufficient for the optimum performance. The real advantage of reservoir computing will be to construct reservoir computers from analog physical systems, but the choice of parameters in physical systems may be restricted, so it may not be possible to operate at the optimum parameter combination.

\subsection{Reservoir Complexity}
The edge of chaos is where a dynamical system is said to have its highest complexity, and therefore its highest computational capacity \cite{packard1988,langton1990,crutchfield1990}. In the examples in \cite{packard1988,langton1990,crutchfield1990}, the dynamical systems cited are all one dimensional maps or cellular automata. For the different nodes types here, arranged in networks to function as reservoir computers, does the complexity increase as the reservoir approaches the edge of chaos?

Entropy is used here as a measure of complexity. With $M$ nodes and $N$ time steps, the entropy for the reservoir is calculated by first mapping the time series from each of the reservoir variables $r_i(j), i=1 \ldots M, j=1 \ldots N$ into symbols using ordinal patterns as described in section \ref{mi_sect}, with a symbol length of 4. If the ordering of points is $[1,3,2,4]$, the symbol is converted to an integer in base 4, $\sigma=1\times 4^0+3 \times 4^1+2 \times 4^2+4 \times 4^3$. The 4 sample window was then slid one time step forward along the time series to find the next symbol.

At the $i$'th time step, the set of symbols for each node is $\Sigma(i) = [\sigma_1(i), \sigma_2(i), \ldots \sigma_M(i)]$. Searching through the $M$ dimensional time series, the number of occurrences  of $\Sigma(i)$ in the rest of the time series is counted. Repeating this procedure, the number of times that each $\Sigma$ occurs is found and the probabilities are calculated. The first symbol found is $\Sigma_1$, the second is $\Sigma_2$, etc. The entropy $H_R$ for the reservoir is then calculated as 
\begin{equation}
\label{res_ent}
{H_R} =  - \sum\limits_{k = 1}^{{N_\Sigma }} {p\left( {{\Sigma _k}} \right)\log \left( {{\Sigma _k}} \right)} 
\end{equation}
where $N_{\Sigma}$ is the total number of symbols.

\begin{figure}[ht]
\centering
\includegraphics[scale=0.8] {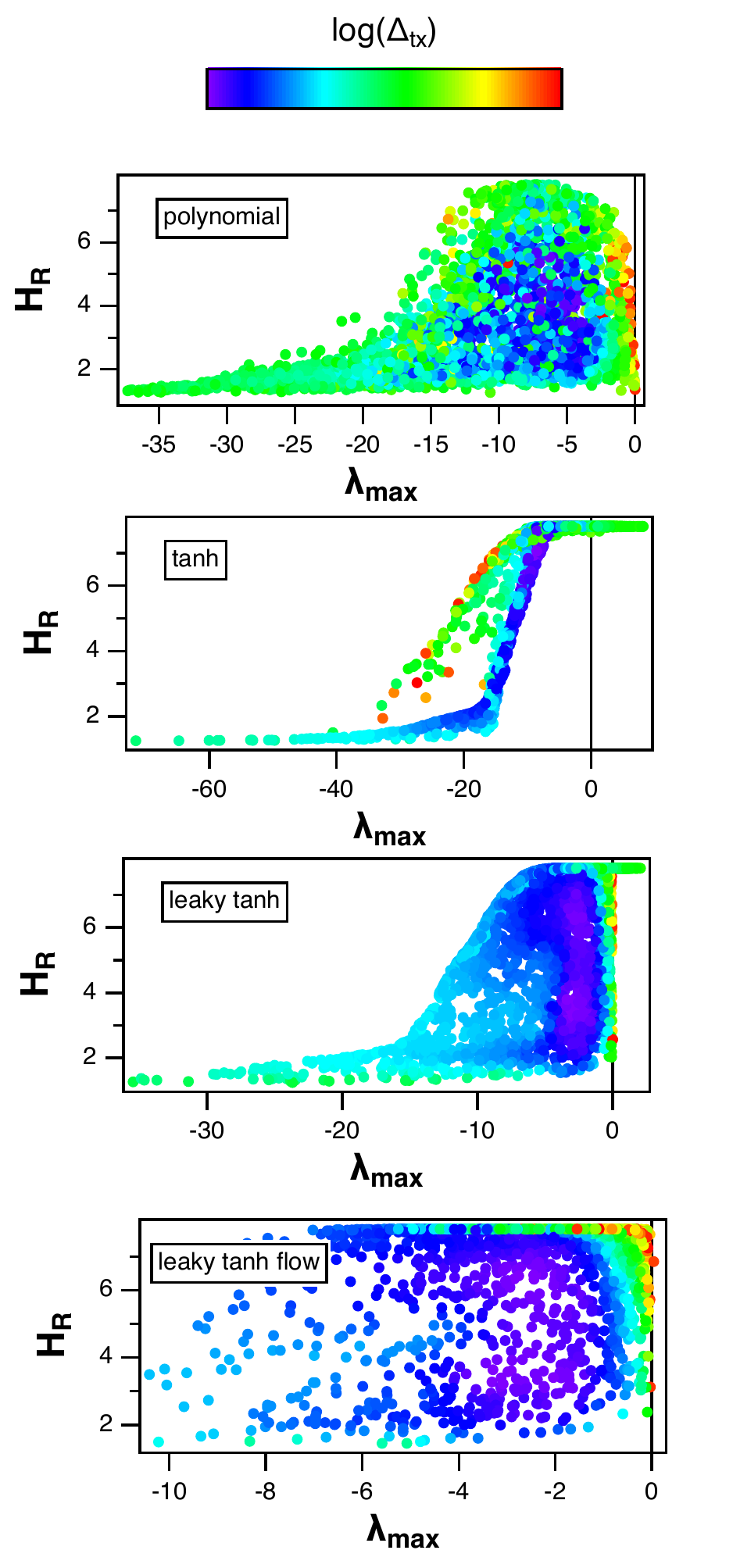} 
  \caption{ \label{entropy_res} Entropy $H_R$ of the reservoir variables $r_i(j), i=1 \ldots M, j=1 \ldots N$ as a function of the largest Lyapunov of the reservoir, $\lambda_{max}$, for four different node types. The points are colored by the log base 10 of the testing error $\Delta_{tx}$, where purple represents the smallest testing error and red represents the largest.}
  \end{figure}

Figure \ref{entropy_res} shows the entropy of the set of reservoir variables $r_i(j), i=1 \ldots M, j=1 \ldots N$ as a function of the largest Lyapunov exponent of the reservoir.

The reservoir entropy for the hyperbolic tangent nodes (tanh) in figure \ref{entropy_res} clearly increases as the maximum Lyapunov exponent approaches the transition from negative to positive, fulfilling the assumptions in the edge of chaos work \cite{packard1988,langton1990,crutchfield1990}. The minimum testing error $\Delta_{tx}$ for the hyperbolic tangent nodes, plotted in figure \ref{random_errors} does come close to the edge of chaos. The hyperbolic tangent is probably the most commonly used node type in reservoir computers, so it is not surprising that conventional wisdom says that the best place to operate a reservoir computer is at the edge of chaos. Still there are also large testing errors near the edge of chaos for the hyperbolic tangent nodes in figure \ref{random_errors}, so simply having complex signals is not sufficient to produce the smallest testing error.

The pattern of reservoir entropy vs. largest Lyapunov exponent is more complicated for the polynomial and leaky hyperbolic tangent map (leaky tanh) nodes. The overall trend is for the reservoir entropy $H_R$ to get larger as the largest Lyapunov exponent $\lambda_{max}$ approaches the transition from negative to positive, but there is a broad range of entropy values near this threshold. In figure \ref{random_errors}, the minimum of the testing error $\Delta_{tx}$ still occurs near the edge of chaos, but once again there may also be very large testing errors near this boundary.

In contrast to the other node types, the reservoir entropy for the leaky hyperbolic tangent flow pictured in figure \ref{entropy_res} does not show any dependance on the largest Lyapunov exponent. In figure \ref{random_errors}, the testing error $\Delta_{tx}$ does go through a minimum as the largest Lyapunov exponent increases, but without a clear definition of what range of Lyapunov exponent constitutes the edge of chaos, it is not possible to say if the minimum testing error occurs at the edge of chaos or not.

\subsection{Fit Signal Entropy}
The fit signal $h(t)$ is a linear combination of node signals, so the fit signal entropy $H_h$ may be different than the reservoir entropy $H_R$.  The fit signal can be more or less complex than the training signal $g(t)$, but the lowest testing error should come when the two signals are equally complex.

\begin{figure}[ht]
\centering
\includegraphics[scale=0.8] {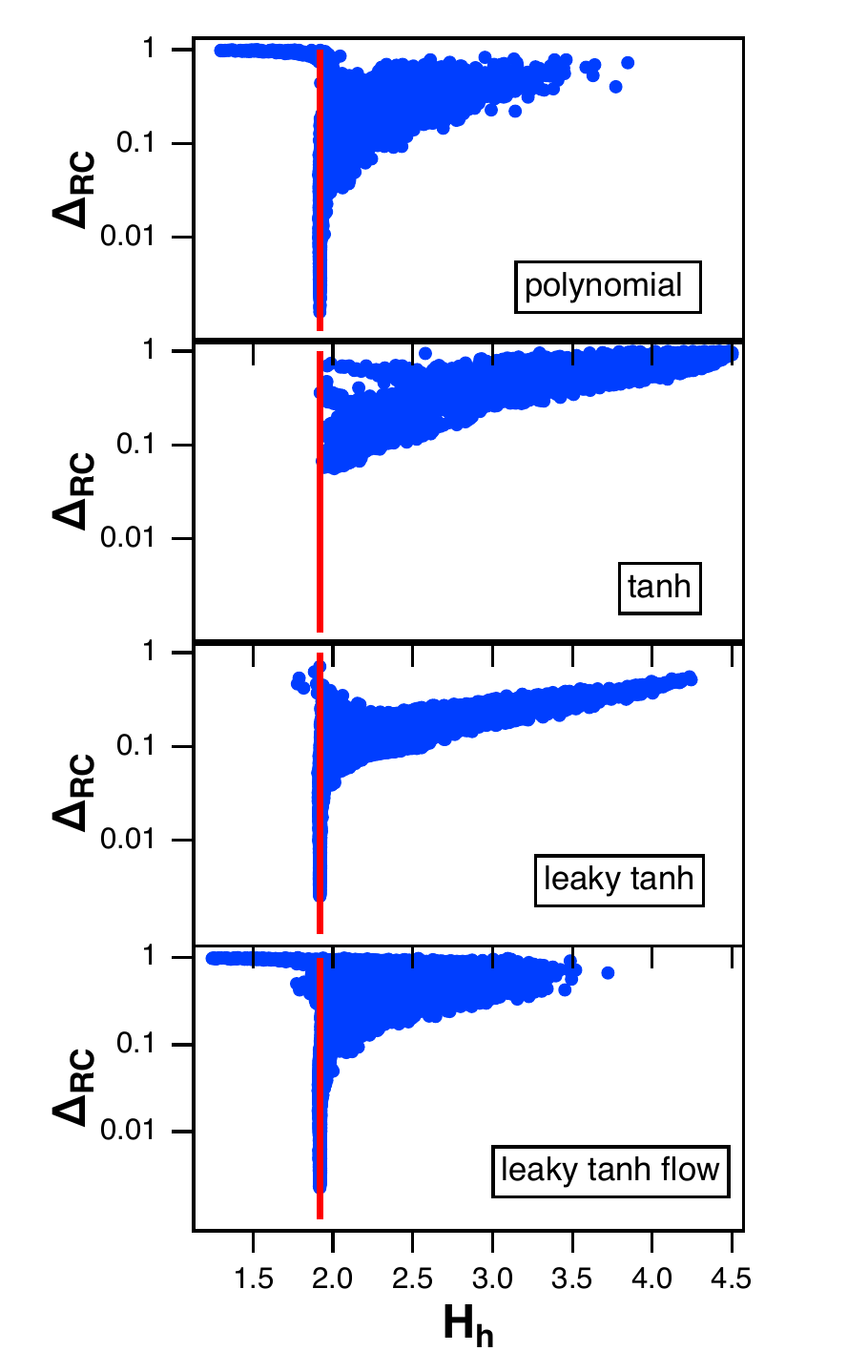} 
  \caption{ \label{err_entropy} The blue dots represent the testing error $\Delta_{tx}$ as a function of the entropy $H_h$ of the fit signal $h(t)$. The red vertical line is the entropy $H_g$ of the training signal $g(t)$. The four node types were the polynomial nodes, the hyperbolic tangent (tanh) nodes, the leaky hyperbolic tangent (leaky tanh) nodes, and the hyperbolic tangent flow (tanh flow) nodes.
  }
  \end{figure}

 Figure \ref{err_entropy} shows how the testing error changed as a function of the entropy of the fit signal, $H_h= -\sum\limits_{j = 1}^{{N_{sh}}} {p\left( {{\sigma _h}\left( j \right)} \right)\log \left[ {p\left( {{\sigma _h}\left( j \right)} \right)} \right]} $. Also plotted in figure \ref{err_entropy} is a vertical red line indicating the entropy of the training signal, or $H_g=-\sum\limits_{i = 1}^{{N_{sg}}} {p\left( {{\sigma _g}\left( i \right)} \right)\log \left[ {p\left( {{\sigma _g}\left( i \right)} \right)} \right]}$. The input signal is the same for all the reservoir computer types. Each plot in figure \ref{err_entropy} was created by taking 6000 random combinations of parameters for each of four different node types, as in section \ref{randpar}. As in that section, the network was not changed.

In all four plots in figure \ref{err_entropy}, the minimum testing error $\Delta_{tx}$ is seen when the entropy of the fit signal, $H_h$, is the same as the entropy of the training signal. If the signals are closely matched, then the entropies should be approximately equal, which is another way of stating that their complexities are equal.

When the entropy of the fit signal is less than or greater than the entropy of the training signal, the testing error $\Delta_{tx}$ is larger. Because the fit signal $h(t)$ is a linear combination of reservoir signals, and the linear combination is different, there is no simple relation between the fit signal entropy and the reservoir entropy. The reservoir entropy does tend to increase towards the edge of chaos for three of the four node types (figure \ref{entropy_res}), but extra complexity alone is not all that is required. Figure \ref{err_entropy} shows that even when the entropies of the fit signal and training signal match, the testing error $\Delta_{tx}$ can be large.

\subsection{Relation to Covariance Rank}
In figures \ref{lorenz_nleq_line}, \ref{tanh_lorenz_line},\ref{umd_lorenz_line} and \ref{tanh_flow_lorenz_line}, the minimum testing error was seen when the covariance rank $\Gamma$ was at a maximum. Increasing the complexity of the reservoir signals $r_i(t)$ led to a larger covariance rank for the reservoir matrix $\Omega$ up to a point. The maximum possible covariance rank for $\Omega$ was equal to the number of nodes, or 100, so further increases in the reservoir signal complexity could not increase the covariance rank past this point. It was shown in \cite{carroll2019} that larger covariance ranks were associated with smaller testing errors, so if increased complexity leads to a larger covariance rank, it decreases the testing error. Once the covariance rank is maximized, however, increasing the complexity of the reservoir signals may not decrease the testing error, and as mentioned in the previous section, the increased complexity might actually increase the testing error.

\section{Conclusions}
It is commonly stated that a reservoir computer works best at the edge of chaos, the point were the maximum Lyapunov exponent for the reservoir goes from negative to positive. The simulations here certainly do not disprove this statement, but they do show that tuning a reservoir computer to the edge of chaos does not guarantee the best performance. Some parameter combinations that leave a reservoir computer near the edge of chaos lead to large testing errors. In a real physical system, the range of parameters is limited, so it may not be possible to tune the system to the optimum parameters. It was also shown that the assumption that underlies the edge of chaos concept does not hold for all reservoir types. The edge of chaos concept was based on well known routes to chaos such as period doubling in one dimensional maps. As the map approached chaos, it exhibited more complex behavior, leading to a larger computational capacity. Reservoir computers are much higher dimensional, and they may not actually become chaotic when the largest Lyapunov exponent becomes positive; some types of reservoir computer may simply become unstable, at which point the reservoir variables will approach $\pm \infty$ (or the power supply voltage). It has been pointed out in \cite{lymburn2019} that the reservoir computer has many Lyapunov exponents, so although the largest Lyapunov exponent is positive, there can be many other negative exponents, so the edge of chaos for such a high dimensional system may not have the same meaning as for a low dimensional system.

The simulations in this paper do show that having a larger mutual information between the fit signal $h(t)$ and the training signal $g(t)$ does lead to better reservoir computer performance. The fit signal $h(t)$ is a linear combination of the time series outputs of the individual reservoir nodes, so the job of the reservoir computer is to create from the input signal $s(t)$ a set of signals that have the largest mutual information with the training signal $g(t)$.

It was also shown that the rank of the reservoir covariance matrix $\Omega^T\Omega$ was important for obtaining a small testing error. The lowest testing errors were seen when the rank of the covariance matrix was at its maximum value. A future question for reservoir computer studies is how do different nonlinear nodes increase this covariance rank?

This work was supported in part by the Office of Naval Research through the Naval Research Laboratory's Basic Research Program.

\section{References}
\bibliography{res_comp_and_MI.bib}{}

\end{document}